# Timing Detection of Eclipsing Binary Planets and Transiting Extrasolar Moons

**Laurance R. Doyle**
SETI Institute, 2035 Landings Drive, Mountain View, CA 94043

**Hans-Jörg Deeg**
Instituto de Astrofisica de Canarias, E-38200 La Laguna, Tenerife, Spain

Abstract

We investigate the improved detection of extrasolar planets around eclipsing binaries using eclipse minima timing and extrasolar moons around transiting planets using transit timing offered by the upcoming COROT (ESA, 2005), Kepler (NASA, 2007), and Eddington (ESA 2008) spacecraft missions. Hundreds of circum-binary planets should be discovered and a thorough survey of moons around transiting planets will be accomplished by these missions.

Eclipsing Binary Minima Timing

In addition to providing a higher probability for edge-on planetary orbital alignments for transit events (see Deeg et al. 1998, Doyle et al. 2000), eclipsing binary star systems also allow the discovery of even non-planar outer giant planets because the eclipse minima act as definitive events that can be timed for offsets around the binary/planet center of mass—the barycenter (Doyle et al. 1996, Deeg et al. 2000). A giant outer planet around any such systems will offset the binary causing a periodic early or late arrival of the light from the eclipse minimum events across the offset barycenter by an amount (Schneider and Doyle 1995):

$$\delta t = M_p a_p / c M_* \qquad (1)$$

where $M_p$ and $M_*$ are the masses of the planet and star, respectively, $a_p$ is the semi-major axis of the planet's orbit, and $c$ is the speed of light.

The precision with which the timing of the binary eclipse minima at time $t_0$ can be determined is a function of the stellar brightness $L(t_i, t_0)$ at any given time $t_i$, so that the relationship between the measurement error in brightness $\delta_L$ and the error in eclipse minimum timing $\delta_{t_0}$ can be determined by the basic considerations of error propagation (Press 1986, for example) and given as:

$$\delta_{t_0} = \delta_L \left[ \sum_i \left( \frac{\partial L(t_i, t_0)}{\partial t_0} \right)^2 \right]^{-\frac{1}{2}} \qquad (2).$$

(This equation corrects a typesetting error in Equation 3 of Doyle et al. 1996.) If the measurement of $L$ is taken at equidistant points $t_i$, where $\Delta t = t_{i+1} - t_i$ the relation $L(t_i, t_0 + \Delta t) = L(t_{i-1}, t_0)$ holds, which allows a very simple numerical calculation of the derivative in Equation 2 from a single lightcurve (measure of star brightness with time), using:

$$\frac{\partial L(t_i, t_0)}{\partial t_0} \approx \frac{L(t_{i-1}, t_0) - L(t_{i+1}, t_0)}{2\Delta t} \qquad (3).$$

An approximation for an easy application to observational data can also be derived through an analytical integration of Equation 1 for an eclipse with a triangular shape (the following equation therefore also sets a strict lower limit):

$$\delta_{t_0} \approx \delta_L \frac{T_{ec}}{2\Delta L \sqrt{N}} \qquad (4),$$

where $T_{ec}$ is the duration of the eclipse event (from first to last contact), $N$ is the number of observational data points taken during $T_{ec}$, and $\Delta L$ is the relative depth of the eclipse (with the out-of-eclipse brightness being unity). As an example, for observations of the CM Draconis eclipsing binary, $T_{ec} \approx 80$ minutes = 4800 seconds, $\Delta L = 0.46$, with a relative photometric error (error in brightness measurements) of $\delta_L = 0.01$, taken with a cadence of 5 seconds ($N = 960$), we obtain an eclipse minimum timing error of about 1.7 seconds or less.

Deeg *et al.* (2000) used this method to set an upper limit of 2-3 Jupiter masses for the presence of third bodies (with periods less than 3 years) around the CM Draconis system. The actual average timing errors in this work were about 6 seconds.

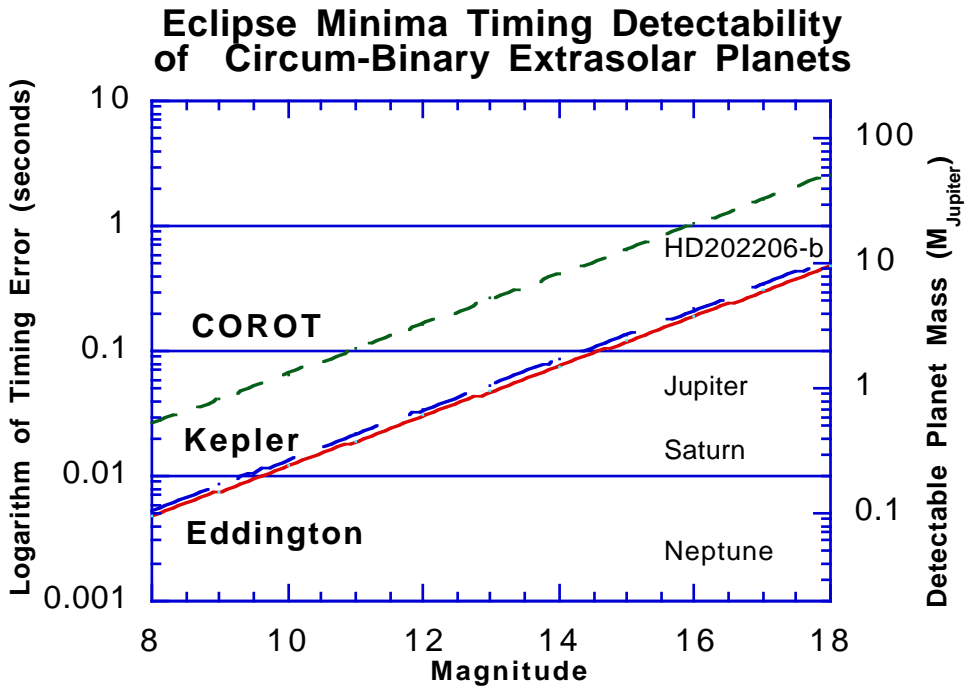

FIGURE 1. Eclipse minima timing precision achievable with the COROT, Kepler, and Eddington spacecraft missions for binary systems consisting of two solar masses, a planet of a given mass at 2/3 AU (140-day period), with 20 eclipse minima having been measured. Note that the planet does not have to be in a coplanar orbit with the eclipsing binary, although the measureable offset will, of course, have a planetary orbital inclination sine dependence. Shown are example planetary masses including HD202206-b, whose minimum mass is closer to that of a brown dwarf.

Spacecraft such as the COROT, Kepler, and Eddington missions can be expected to achieve a much higher photometric precision (at least a few parts in $10^5$) as well as many more eclipse minima over the 3-to-4-years of continuous monitoring of a given crowded stellar field (for example, Borucki *et al.* 1997). As an example, assuming a total eclipsing binary mass of two solar masses, a planetary orbital period of about 140 days (i.e. a semi-major axis of about 2/3 of an AU), with 20 eclipse minima recorded, and a sampling rate of 15 minutes, better than 0.1-second precision in the timing of the eclipse minima should be achievable, as shown in Figure 1 (where photon noise dominates so that the precision is proportional to the effective mirror size of each spacecraft.) Hundreds of new outer giant planets around eclipsing binaries should be discoverable by these missions, then, using this approach to the same photometric data sets they are already going to be acquiring.

Transiting Planet Timing

The precise timing of planetary transits themselves (hundreds of which are expected to be discovered by the COROT, Kepler, and Eddington missions) can reveal the presence of extrasolar moons around these planets, as first pointed out by Sartoretti and Schneider (1999).

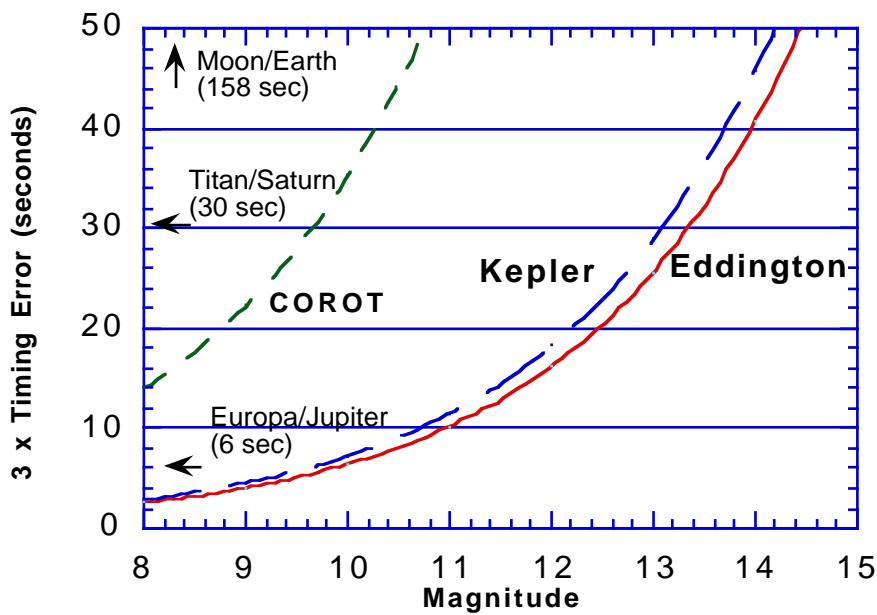

FIGURE 2. For the hundreds of transiting planets that the COROT, Kepler, and Eddington missions will discover, timing can also distinguish if any of these planets have moons in orbit around them. An Earth/Moon-type system is easily distinguished. But, interestingly, Europa/Jupiter-type systems would also be discoverable. Note that the vertical axis for the expected spacecraft timing errors are three-sigma values.

Again, the timing offset of the planet produced by the presence of a moon around the planet/moon barycenter may be formulated (after Sartoretti and Schneider 1999) as:

$$\delta t \approx \frac{a_m M_m P_p}{\pi a_p M_p} \quad (5),$$

where $a_m$ and $M_m$ are the semi-major axis and the mass of the extrasolar moon, respectively, and $P_p$ is the orbital period of the planet. Using a similar formulation to estimate the timing errors, we arrive at the detection limits given in Figure 2. An Earth-Moon system is easily detectable (once the precision needed to detect an Earth-sized planet in transit has been achieved). But even a Europa/Jupiter configuration would be within the limits of detectability for the Kepler and Eddington Missions. In perhaps most cases the accompanying moon would not show up in transit so that the timing offsets of the transiting planet itself would be the only way to detect their presence. From orbital stability considerations no detectably massive moons may be expected for very short period (less than a week) giant planets, however (Barnes and O'Brien 2002). Since, in this case, the offset of a transiting planet by its moon is more of an astrometric effect (the spacial displacement of the planet by the moon against the stellar disc) rather than a displacement toward of away from the observer, this effect is essentially independent of the inclination to the line-of-sight of the displacing moon's orbit.

## Conclusions

Within the past decade over 100 extrasolar planetary systems have been discovered but, as of this date, none have been found in circum-binary orbits. The eclipsing binary timing method may be the only method that can achieve such detections (besides possibly the gravitational microlensing technique) and this is of some importance since close binary systems might be considered of particular interest in understanding planet formation processes because of their different, and likely more complex, angular momentum histories. Discovering the prevalence of moons around extrasolar planets will also be of great interest both for understanding the formation of our own moon, but also—since Europa/Jupiter-type systems are also detectable by both the Kepler and Eddington missions—as possible tidally heated habitable zones that might be similar to the ocean environment that exists under the ice crust on Jupiter's moon Europa today. The rough statistical prevalence of circum-binary planets and extrasolar moons are questions that we should finally begin to have answers to within the next exciting decade.